\documentclass[conference]{IEEEtran}
\IEEEoverridecommandlockouts
\usepackage{cite}
\usepackage[a4paper, left=1.27cm, right=1.2cm, top=1.9cm, bottom=2.9cm]{geometry}
\usepackage{amsmath,amssymb,amsfonts}
\usepackage[bookmarks=false]{hyperref}
\hypersetup{draft}
\usepackage{algorithm}
\usepackage[a-1b]{pdfx}
\usepackage{algpseudocode}
\usepackage{lettrine}
\usepackage{graphicx}
\usepackage{textcomp}
\usepackage{xcolor}
\def\BibTeX{{\rm B\kern-.05em{\sc i\kern-.025em b}\kern-.08em
    T\kern-.1667em\lower.7ex\hbox{E}\kern-.125emX}}
\begin{document}

\title{TeraRIS NOMA-MIMO Communications for 6G and Beyond Industrial Networks\\
\thanks{The contributions of Prof. Markku Juntti and Dr. Nhan Thanh Nguyen from the Centre for Wireless Communications, University of Oulu, Finland, are gratefully acknowledged in this work.}
}

\author{
    \IEEEauthorblockN{Ali Raza\IEEEauthorrefmark{1}, Muhammad Farhan Khan\IEEEauthorrefmark{2},  Zeeshan Alam\IEEEauthorrefmark{3}, 
    Muhammad Saad\IEEEauthorrefmark{4},
    Ilyas Saleem\IEEEauthorrefmark{5}, \\
    Muhammad Ahmed Mohsin\IEEEauthorrefmark{6}, Muhammad Ali Jamshed\IEEEauthorrefmark{7}
   }
    \IEEEauthorblockA{\IEEEauthorrefmark{1}Centre for Wireless Communications, University of Oulu, P.O.Box 4500, FI-90014, Finland\\}
    \IEEEauthorblockA{\IEEEauthorrefmark{2}School of Computer Science and Information Technology, University College Cork, Cork, Ireland \\}
    \IEEEauthorblockA{\IEEEauthorrefmark{3}Faculty of Computer Science,  University of New Brunswick,
NB, Canada\\}  
    \IEEEauthorblockA{\IEEEauthorrefmark{4}SEECS, National University of Science and Technology, Islamabad, Pakistan\\}    \IEEEauthorblockA{\IEEEauthorrefmark{5}Faculty of Science and Engineering, Macquarie University, Sydney, NSW, Australia \\}
    \IEEEauthorblockA{\IEEEauthorrefmark{6}Department of Electrical Engineering, Stanford University, Stanford, CA,  94305, USA \\}
    \IEEEauthorblockA{\IEEEauthorrefmark{7}College of Science and Engineering, University of Glasgow, UK}
    
    \IEEEauthorblockA{Email:  ali.raza@oulu.fi, Farhan.khan@cs.ucc.ie, Muhammad.alam@unb.ca }}

\maketitle

\begin{abstract}
This paper presents a joint framework that integrates reconfigurable intelligent surfaces (RISs) with Terahertz (THz) communications and non-orthogonal multiple access (NOMA) to enhance smart industrial communications. The proposed system leverages the advantages of RIS and THz bands to improve spectral efficiency, coverage, and reliability—key requirements for industrial automation and real-time communications in future 6G networks and beyond. Within this framework, two power allocation strategies are investigated: the first optimally distributes power between near and far industrial nodes, and the second prioritizes network demands to enhance system performance further. A performance evaluation is conducted to compare the sum rate and outage probability against a fixed power allocation scheme. Our scheme achieves up to a $23$\% sum-rate gain over fixed PA at $30$ dBm. Simulation results validate the theoretical analysis, demonstrating the effectiveness and robustness of the RIS-assisted NOMA-MIMO framework for THz-enabled industrial communications.
\end{abstract}


\section{Introduction}

\begin{figure*}[t]
    \centering
    \includegraphics[width=0.5\textwidth]{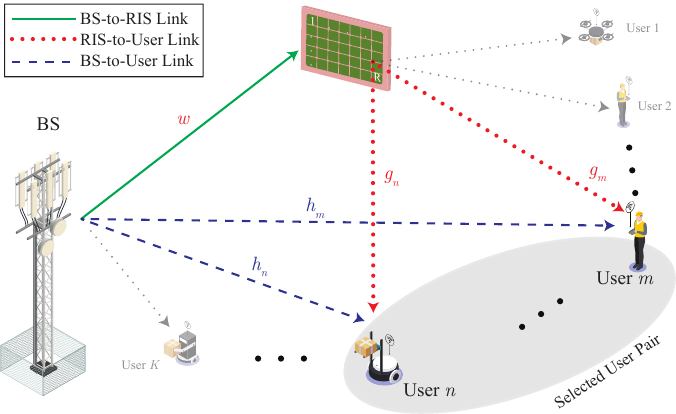}
    \caption{System model configuration for considered TeraRIS NOMA-MIMO industrial communications framework.}
    \label{image2}
\end{figure*}

Motivated by the potential benefits of THz, RIS, MIMO, and NOMA, it is necessary to integrate these technologies to improve bandwidth, energy efficiency, network coverage, and connectivity at higher data rates~\cite{10464446, 10462124}. In this paper, we consider a joint RIS-assisted NOMA-MIMO system operating at the THz band, which we referred to as TeraRIS NOMA-MIMO system for industrial communications. This framework offers a controllable propagation environment, enabling advanced futuristic applications for Industry 5.0 and beyond. The main contributions of this work include: First, we propose a power allocation algorithm for SISO-NOMA, extended from \cite{35} to accommodate MIMO systems with RIS and THz band characteristics. Second, we perform a comparative performance analysis of fixed power allocation (PA) and fair PA schemes is conducted, evaluating key performance indicators such as sum rate and outage probability.

\section{ System Model}

We consider a wireless communication system in Fig.~\ref{image2} within a critical industrial setting with a total number of $K$ users. The number of antennas at the base station (BS) is indicated by $N$, and the number of antennas at user $k$ is represented by $M_k$. The system employs an RIS consisting of $R$ reflecting elements. The transmitted signal from the BS is denoted by the $N\times1$ vector $\boldsymbol{\mathrm{x}}_k$. The channel between BS and user $k$ via the RIS is influenced by both the BS-to-RIS and RIS-to-user $k$ channels, denoted as $\boldsymbol{w}$ and $\boldsymbol{g}_k$, respectively (as shown in Fig.~\ref{image2}). This combined channel is represented by the $M_k\times N$ matrix ${\boldsymbol{\mathrm{G}}}_{{\mathrm{RIS}}}^k$. Furthermore, the direct channel between the BS and user $k$ is denoted by $M_k\times N$ matrix ${\boldsymbol{\mathrm{H}}}_{{\mathrm{D}}}^k$. The elements of ${\boldsymbol{\mathrm{G}}}_{{\mathrm{RIS}}}^k$ and ${\boldsymbol{\mathrm{H}}}_{{\mathrm{D}}}^k$, denoted by $g_{ij}^k$ and $h_{ij}^k$, represent the channel gain between the $i$-th transmit antenna and the $j$-th receive antenna. The received signal at user $k$, $\boldsymbol{\mathrm{y}}_k$, is an $M_k\times 1$ vector corrupted by additive white Gaussian noise (AWGN) $\boldsymbol{\mathrm{n}}_k$, which is also an $M_k\times 1$ vector.

\subsection{Channel Model}

\begin{figure}[t]
    \centering
    \includegraphics[width=0.6\columnwidth]{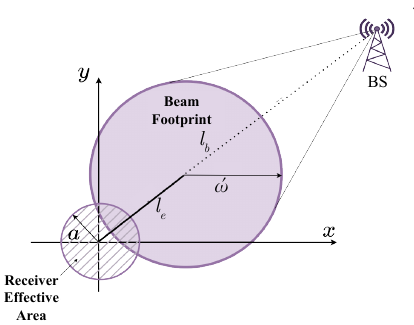}
    \caption{Illustration of beam alignment error, depicting the effects of beamwidth and receiver misalignment in THz communications.}
    \label{image3}
\end{figure}

The considered framework focuses on a two-user NOMA pair near-user $n$ and far-user $m$ to simplify the analysis. This model facilitates deriving SINR and capacity expressions while providing insights into NOMA principles and evaluating the proposed power allocation algorithms \cite{36}. In the considered framework, a multi-ray model with $q=0,\dots, Q$ rays is used \cite{41, }. The 
first ray represents the line-of-sight (LoS) path, while the remaining ones correspond to single-bounce reflected non-line-of-sight (NLoS) channels.\footnote{Due to the high path loss of THz signals caused by scattering and absorption, only single-bounce NLoS paths are considered.} The channel attenuation of the LoS path between the BS and user $k$ is given as follows \cite{43}
\begin{equation}\resizebox{0.85\columnwidth}{!}{$
h^k_{ij}\left(f\right)\mathrm{=}\underbrace{\frac{c}{\mathrm{4}\pi fd^k_{ij}}}_{\mathrm{:=}{\mathrm{\Delta }}_1}\underbrace{\mathrm{exp}\left[-\frac{\kappa \left(f\right)}{\mathrm{2}}d^k_{ij}\right]}_{\mathrm{:=}{\mathrm{\Delta }}_2}\underbrace{{\left({\mathrm{erf} \left(u\right)\ }\right)}^{\mathrm{2}}{\mathrm{exp} (-\frac{2{l_e}^2}{{w_{e}}^2}\ })}_{\mathrm{:=}{\mathrm{\Delta }}_3}.
\label{GrindEQ2}
$}\end{equation}
In \eqref{GrindEQ2}, ${\mathrm{\Delta }}_1$ and ${\mathrm{\Delta }}_2$ denote path and absorption losses, respectively, assuming unity antenna gains. The absorption coefficient $\kappa(f)$ is defined at the central frequency $f$, while $d^k_{ij}$ represents the distance between the $i$-th BS antenna and the $j$-th user antenna. Molecular absorption is typically calculated using radiative transfer theory and the HITRAN database, ensuring theoretical accuracy grounded in experimental data \cite{45}. The term ${\mathrm{\Delta }}_3$ accounts for misalignment fading, a critical issue in THz communications due to narrow beamwidths. As shown in Fig.~\ref{image3}, a receiver with aperture radius $a$ receives a symmetric beam with beamwidth $\acute{\omega }$ transmitted over distance $l_b$, with $l_e$ denoting a pointing error. The resulting misalignment gain is approximated by ${\mathrm{\Delta }}_3$, where $w_{e} = \sqrt{ {\acute{\omega }}^2\frac{\sqrt{\pi }\mathrm{erf}\mathrm{}\left(u\right)}{2u{\mathrm{exp} \left(-u^2\right)\ }}}$. The function $\mathrm{erf}(\cdot)$ is an error function and is evaluated as described in \cite{40} and $u=\frac{\sqrt{\pi }a}{\sqrt{2}\acute{\omega }}$.

For the $q$-th NLOS path, let ${\beta }_q(f)$ denote the path loss factor and ${\tau }_q$ the delay spread relative to the LoS path. The BS-to-user $k$ channel response is given by \cite{41}
\begin{equation}\resizebox{0.85\columnwidth}{!}{$
h^k_{ij}\mathrm{=}h^k_{ij}\left(f\right)\left(\mathrm{1}+\sqrt{\frac{\mathrm{1}}{Q-1}}\sum^{Q-1}_{q\mathrm{=1}}{}{\beta }_q\left(f\right){\mathrm{exp} \left[-\mathrm{2}\pi f{\tau }_q\right]\ }\right).
\label{GrindEQ5}
$}\end{equation}
Now, the matrix for the MIMO channel from BS to direct user $k$ for the considered framework can be denoted as follows
\begin{equation}\resizebox{0.6\columnwidth}{!}{$
{\boldsymbol{\mathrm{H}}}_{{\mathrm{D}}}^k={\boldsymbol{\mathrm{H}}}^k_{ij}=
    \begin{bmatrix}
    h^k_{11} & h^k_{12} & \cdots  & h^k_{1N} \\ 
    h^k_{21} & h^k_{22} & \cdots  & h^k_{2N} \\ 
    \vdots  & \vdots  & \ddots  & \vdots  \\ 
    h^k_{M_11} & h^k_{M_22} & \cdots  & h^k_{M_kN}
    \end{bmatrix}
\label{GrindEQ6}.
$}\end{equation}
The channel gain of the signal from the BS to the user $k$ reflected from the RIS can be obtained as presented in \cite{46} in the following manner
\begin{equation}\resizebox{0.85\columnwidth}{!}{$
g^k_{ij}\mathrm{=}\left(\frac{{\eta }_r\mathrm{exp[}j{\phi }_r\mathrm{]}\lambda }{\mathrm{8}\sqrt{{\pi }^{\mathrm{3}}}r_{ir}r_{rj}}\right)\mathrm{exp}\left[\frac{-\kappa \mathrm{(}f\mathrm{)}d^k_{ij}}{\mathrm{2}}\right]\mathrm{exp}\left[\frac{-j\mathrm{2}\pi \mathrm{(}r_{ir}+r_{rj}\mathrm{)}}{\lambda }\right],
\label{GrindEQ7}
$}\end{equation}
where the phase shift ($\phi_r$) and reflection coefficient ($\eta_r$) of the $r$-th RIS element are assumed to be known. Phase shift optimization for RIS is beyond the scope of this paper. In \eqref{GrindEQ7}, $\lambda = c/f$ is the wavelength, $r_{ir}$ is the distance from the $i$-th BS antenna to the $r$-th RIS element, and $r_{rj}$ is the distance from the $r$-th RIS element to the $j$-th user antenna. The MIMO channel matrix from BS to user $k$ reflected by RIS can be given as
\begin{equation}\resizebox{0.6\columnwidth}{!}{$
{\boldsymbol{\mathrm{G}}}_{{\mathrm{RIS}}}^k={\boldsymbol{\mathrm{G}}}^k_{ij}=
    \begin{bmatrix}
    g^k_{11} & g^k_{12} & \cdots  & g^k_{1N} \\ 
    g^k_{21} & g^k_{22} & \cdots  & g^k_{2N} \\ 
    \vdots  & \vdots  & \ddots  & \vdots  \\ 
    g^k_{M_11} & g^k_{M_21} & \cdots  & g^k_{M_kN}
    \end{bmatrix},
\label{GrindEQ8}
$}\end{equation}
and the overall channel between the BS and user $k$ can be written as
\begin{equation}\resizebox{0.29\columnwidth}{!}{$
{\boldsymbol{\mathrm{H}}}_k{\mathrm{=\ }}{\boldsymbol{\mathrm{H}}}_{{\mathrm{D}}}^k {+} {\boldsymbol{\mathrm{G}}}_{{\mathrm{RIS}}}^k.
\label{GrindEQ9}
$}\end{equation}

\subsection{Received Signal and SINR}
A signal $\boldsymbol{x}$ of dimension $(N\times 1)$ comprising the superimposed information of users $1$ to $K$ is transmitted from the BS, which can be expressed as follows
\begin{equation}\resizebox{0.4\columnwidth}{!}{$
\boldsymbol{\mathrm{x}}=\sum^K_{k=1}{\sqrt{p_{\circ }{\alpha }_k}}x_k,
\label{GrindEQ10}
$}\end{equation}
where ${\alpha}_k$ is the power allocation coefficient for user $k$, with $\sum^K_{k=1}{{\alpha }_k}=1$, $p_{\circ }$ is the transmitted power, and $x_k$ is the message signal for user $k$. The received signal at user $k$ can be expressed as follows 
\begin{equation}\resizebox{0.88\columnwidth}{!}{$
\displaystyle
\begin{array}{lll}
{\boldsymbol{\mathrm{y}}}_k &=& {\boldsymbol{\mathrm{H}}}_k \boldsymbol{\mathrm{x}}+{\boldsymbol{\mathrm{n}}}_k,\\
 &=& {\boldsymbol{\mathrm{H}}}_k\sum^K_{k=1}{\sqrt{p_{\circ }{\alpha }_k}}x_k+{\boldsymbol{\mathrm{n}}}_k,\\
 &=& {\boldsymbol{\mathrm{H}}}_k\sqrt{p_{\circ }{\alpha }_k}x_k+\sum^K_{k\neq j}{{\boldsymbol{\mathrm{(}}\boldsymbol{\mathrm{H}}}_k\sqrt{p_{\circ }{\alpha }_j}}x_j)+{\boldsymbol{\mathrm{n}}}_k.
\end{array}
\label{received}
$}\end{equation}
In a multi-user NOMA environment, user $n$ successively decodes user $m$'s message using successive interference cancellation (SIC) from $m=1$ to $m=n-1$. In \eqref{received}, ${\boldsymbol{\mathrm{n}}}_k\mathrm{\sim }\mathcal{CN}(0,{\sigma^2_\circ\boldsymbol{I}})$ is the AWGN and $\sigma^2_\circ$ is the variance of the noise. The received signal at user $n$ for decoding user $m$'s message is expressed as follows
\begin{equation}\resizebox{0.89\columnwidth}{!}{$
{\boldsymbol{\mathrm{y}}}_{n\to m}=\sqrt{p_{\circ }{\alpha }_m}x_m{\boldsymbol{\mathrm{H}}}_n+\sum^K_{l=m+1}{(\sqrt{p_{\circ }{\alpha }_l}}x_l){\boldsymbol{\mathrm{H}}}_n+{\boldsymbol{\mathrm{n}}}_n.
\label{GrindEQ12}
$}\end{equation}
The received signal at user $n$, for decoding its own message, is expressed as
\begin{equation}\resizebox{0.85\columnwidth}{!}{$
{\boldsymbol{\mathrm{y}}}_n=\sqrt{p_{\circ }{\alpha }_n}x_n{\boldsymbol{\mathrm{H}}}_n+\sum^K_{l=n+1}{(\sqrt{p_{\circ }{\alpha }_l}}x_l){\boldsymbol{\mathrm{H}}}_n+{\boldsymbol{\mathrm{n}}}_n.
\label{GrindEQ13}
$}\end{equation}
where the first term inside the bracket represents user $n$'s self-signal, and the second term is the interference from higher-order terms (user $n+1$ to user $K$). In \eqref{GrindEQ13}, ${\boldsymbol{\mathrm{n}}}_n\mathrm{\sim }\mathcal{CN}(0,{\sigma^2_\circ\boldsymbol{I}})$ is AWGN. The received signal at user $m$ to decode its own message is followed by
\begin{equation}\resizebox{0.85\columnwidth}{!}{$
{\boldsymbol{\mathrm{y}}}_m=\sqrt{p_{\circ }{\alpha }_m}x_m{\boldsymbol{\mathrm{H}}}_m+\sum^K_{l=m+1}{(\sqrt{p_{\circ }{\alpha }_l}}x_l){\boldsymbol{\mathrm{H}}}_m+{\boldsymbol{\mathrm{n}}}_m,
\label{GrindEQ14}
$}\end{equation}
where the first term inside the bracket represents user $m$'s self-signal, and the second term is the interference from higher-order terms (user $m+1$ to user $K$), and ${\boldsymbol{\mathrm{n}}}_m\mathrm{\sim }\mathcal{CN}(0,{\sigma^2_\circ\boldsymbol{I}})$ is AWGN.

To simplify and facilitate comprehension, we consider the ideal SIC detection for each user. The SINR for user $n$ to decode the message of user $m$ is given as
\begin{equation}\resizebox{0.6\columnwidth}{!}{$
{\zeta }_{n\to m}=\frac{{p_{\circ }{\alpha }_m\left\|{\boldsymbol{\mathrm{H}}}_n\right\|}^2}{{p_{\circ }\left\|{\boldsymbol{\mathrm{H}}}_n\right\|}^2\sum^K_{l=m+1}{{\alpha }_l}+\sigma^2_\circ\boldsymbol{I}},
\label{GrindEQ15}
$}\end{equation}
where the operation $\parallel \cdot \parallel$ denotes the Frobenius norm or another appropriate norm for vectors or matrices, given as $\parallel \boldsymbol{\mathrm{A}}\parallel=\sqrt{trace(\boldsymbol{\mathrm{A}}{\boldsymbol{\mathrm{A}}}^{\mathrm{H}})}$, where ${\left(\cdot \right)}^{\mathrm{H}}$ is the conjugate transpose of any matrix. From \eqref{GrindEQ13}, the SINR for user $n\ (n<K)$ is given as
\begin{equation}\resizebox{0.5\columnwidth}{!}{$
{\zeta }_n=\frac{{p_{\circ }{\alpha }_n\left\|{\boldsymbol{\mathrm{H}}}_n\right\|}^2}{{p_{\circ }\left\|{\boldsymbol{\mathrm{H}}}_n\right\|}^2\sum^K_{l=n+1}{{\alpha }_l}+{\sigma }^2_{\circ }\boldsymbol{I}},
\label{GrindEQ16}
$}\end{equation}
and the SINR for user $m$, to decode its own message and considering all other messages as interference can be given as follows
\begin{equation}\resizebox{0.5\columnwidth}{!}{$
{\zeta }_m=\frac{{p_{\circ }{\alpha }_m\left\|{\boldsymbol{\mathrm{H}}}_m\right\|}^2}{{p_{\circ }\left\|{\boldsymbol{\mathrm{H}}}_m\right\|}^2\sum^K_{l=m+1}{{\alpha }_l}+{\sigma }^2_{\circ }\boldsymbol{I}}.
\label{GrindEQ17}
$}\end{equation}

\section{Capacity and Outage of TeraRIS NOMA-MIMO Communications System}

Using the definition of capacity and the SINR given above, the capacity of user $n\ (n<K)$ can be expressed as follows
\begin{equation}\resizebox{0.6\columnwidth}{!}{$
C_n={{\mathrm{log}}_2 \left(1+\frac{{p_{\circ }{\alpha }_n\left\|{\boldsymbol{\mathrm{H}}}_n\right\|}^2}{{p_{\circ }\left\|{\boldsymbol{\mathrm{H}}}_n\right\|}^2\sum^K_{l=n+1}{{\alpha }_l}+{\sigma }^2_{\circ }\boldsymbol{I}}\right)\ },
\label{GrindEQ18}
$}\end{equation}
and the capacity of user $n$ while decoding user $m$'s message is given as
\begin{equation}\resizebox{0.7\columnwidth}{!}{$
C_{n\to m}={{\mathrm{log}}_2 \left(1+\frac{{p_{\circ }{\alpha }_m\left\|{\boldsymbol{\mathrm{H}}}_n\right\|}^2}{{p_{\circ }\left\|{\boldsymbol{\mathrm{H}}}_n\right\|}^2\sum^K_{l=m+1}{{\alpha }_l}+{\sigma }^2_{\circ }\boldsymbol{I}}\right).\ }
\label{GrindEQ19}
$}\end{equation}
However, the capacity of user $m$ while decoding the user its own message is given as
\begin{equation}\resizebox{0.7\columnwidth}{!}{$
C_m={{\mathrm{log}}_2 \left(1+\frac{{p_{\circ }{\alpha }_m\left\|{\boldsymbol{\mathrm{H}}}_m\right\|}^2}{{p_{\circ }\left\|{\boldsymbol{\mathrm{H}}}_m\right\|}^2\sum^K_{l=m+1}{{\alpha }_l}+{\sigma }^2_{\circ }\boldsymbol{I}}\right).\ }
\label{GrindEQ20}
$}\end{equation}
By employing SIC, the user $K$ with the highest-order effectively mitigates interference caused by the lower-order users. Thus, the capacity is reproduced as
\begin{equation}\resizebox{0.5\columnwidth}{!}{$
C_K={{\mathrm{log}}_2 \left(1+\frac{{p_{\circ }{\alpha }_K\left\|{\boldsymbol{\mathrm{H}}}_K\right\|}^2}{\sigma^2_\circ \boldsymbol{I}}\right)\ }.
\label{GrindEQ21}
$}\end{equation}

The probability density function (PDF) of the envelopes received for the proposed framework can be expressed as follows
\begin{equation}\resizebox{0.6\columnwidth}{!}{$
f_{\mathrm{(}\cdot \mathrm{)}}\mathrm{(}x\mathrm{)=}\frac{2m^{m_{\mathrm{(}\cdot \mathrm{)}}}_{\mathrm{(}\cdot \mathrm{)}}x^{2m_{\mathrm{(}\cdot \mathrm{)}}-\mathrm{1}}}{\mathrm{\Gamma }\mathrm{(}m_{\mathrm{(}\cdot \mathrm{)}}\mathrm{)}}\mathrm{exp[}-m_{\mathrm{(}\cdot \mathrm{)}}x^2\mathrm{],}
\label{GrindEQ1}
$}\end{equation}
where $m_{\mathrm{(}\cdot \mathrm{)}}$ represents the fading shape parameter, and $\mathrm{\Gamma }\mathrm{(}\cdot \mathrm{)}$ denotes the Gamma function, $\mathrm{\Gamma }\ (z)\ =\int^{\infty }_0{t^{z-1}e^{-t}\ dt}$ for $\mathfrak{R}(z)>0$ \cite{40}. By using the same convention used in \cite{al2022ergodic}, which includes partial fraction expansion and residue theory-based integration simplification, the closed-form equations for capacities can be expressed in the  following, so, $C_{n\to m}$ is given by
\begin{equation}\resizebox{0.85\columnwidth}{!}{$
C_{n\to m}=\boldsymbol{\mathrm{E}}\left[{{\mathrm{log}}_2 \left(1+\frac{{\left\|{\boldsymbol{h}}_n\right\|}^2p_{\circ }{\alpha }_m
}{{p_{\circ }\left\|{\boldsymbol{h}}_n\right\|}^2\sum^K_{l=m+1}{{\alpha }_l}\boldsymbol{I}+{\sigma }^2_{\circ }\boldsymbol{I}}\right)\ }\right],
\label{capn2m1}
$}\end{equation}
by using whitening transformation\textbf{ }${\boldsymbol{h}}_n\triangleq vec\left({\boldsymbol{\mathrm{H}}}_n\right)\ $ such that ${\boldsymbol{h}}_n\sim CN(0,\tilde{\boldsymbol{R}}_n)$. Transforming ${\boldsymbol{h}}_n={\tilde{\boldsymbol{R}}^{\mathrm{H}/2}_n 
\overline{\boldsymbol{h}}}_n$ by using the same approach used in \cite{al2022ergodic} we have
\begin{equation}\resizebox{0.7\columnwidth}{!}{$
C_{n\to m}=\boldsymbol{\mathrm{E}}\left[{{\mathrm{log}}_2 \left(1+\frac{{A'_n\left\|{\overline{\boldsymbol{h}}}_n\right\|}^2}{{B'_n\left\|{\overline{\boldsymbol{h}}}_n\right\|}^2+{\sigma }^2_{\circ }\boldsymbol{I}}\right)\ }\right],
\label{capn2m2}
$}\end{equation}
where $A'_n\mathrm{=}p_{\mathrm{\circ }}{\alpha }_m\tilde{\boldsymbol{R}}^\mathrm{H}_n$ and $B'_n=p_{\mathrm{\circ }}\tilde{\boldsymbol{R}}^\mathrm{H}_n\sum^K_{l=m\mathrm{+1}}{{\alpha }_l}$. By evaluating the $\boldsymbol{\mathrm{E}}[\cdot]$ operation as outlined in \cite{al2022ergodic}, we can express \eqref{capn2m2} as follows
\begin{equation}\resizebox{0.85\columnwidth}{!}{$
\begin{split}
\displaystyle
C_{n\to m} &= \frac{\mathrm{1}}{{\mathrm{ln} (2)\ }}\left[\sum^{NM_k}_{i=1}{\frac{{\lambda }^{NM_k-1}_ie^{\frac{1}{{\lambda }_i}}}{\prod^{NM_k}_{j=1,j\neq i}{\left({\lambda }_i-{\lambda }_j\right)}}} E_1\left(\frac{1}{{\lambda }_i}\right)u\left({\lambda }_i\right) \right. \vspace{5pt} \\ \ \\
& \quad \left. -\sum^{NM_k}_{i=1}{\frac{{\lambda }^{NM_k-1}_ie^{\frac{1}{v_i}}}{\prod^{NM_k}_{j=1,j\neq i}{\left(v_i-v_j\right)}}}E_1\left(\frac{1}{v_i}\right)u\left(v_i\right)\right]
\label{capn2mclosed}
\end{split}
$}\end{equation}
where ${\lambda }_i$ are the eigenvalues of the $A'_n+B'_n$ and $v_i$ are the eigenvalues of $B'_n$.  Moreover, $u(\cdot)$, and $E_1(\cdot)$ denote the unit step and exponential integral functions, respectively.

Now using the same approach used in \eqref{capn2m1}, \eqref{capn2m2}, and \eqref{capn2mclosed} we can rewrite the \eqref{GrindEQ20} in closed-form as follows
\begin{equation}\resizebox{0.85\columnwidth}{!}{$
\begin{split}
\displaystyle
C_m &=\frac{\mathrm{1}}{{\mathrm{ln} (2)\ }}\left[\sum^{NM_k}_{i=1}{\frac{{\lambda }^{NM_k-1}_ie^{\frac{1}{{\lambda }'_i}}}{\prod^{NM_k}_{j=1,j\neq i}{\left({\lambda }'_i-{\lambda }'_j\right)}}}E_1\left(\frac{1}{{\lambda }'_i}\right)u\left({\lambda }'_i\right) \right. 
\vspace{5pt} \\ \ \\
& \quad \left. -\sum^{NM_k}_{i=1}{\frac{{{(\lambda }'_i)}^{NM_k-1}e^{\frac{1}{v'_i}}}{\prod^{NM_k}_{j=1,j\neq i}{\left(v'_i-v'_j\right)}}}E_1\left(\frac{1}{v'_i}\right)u\left(v'_i\right)\right]
\label{capmclosed}
\end{split}
$}\end{equation}
where, ${\lambda }'_i$ are the eigenvalues of the $A''_n+B''_n$, and $v'_i$ are the eigenvalues of $B''_n$. Using the same approach, we can also find $C_n$.

\begin{figure*}[t]
    \centering
    \includegraphics[width=0.9\textwidth]{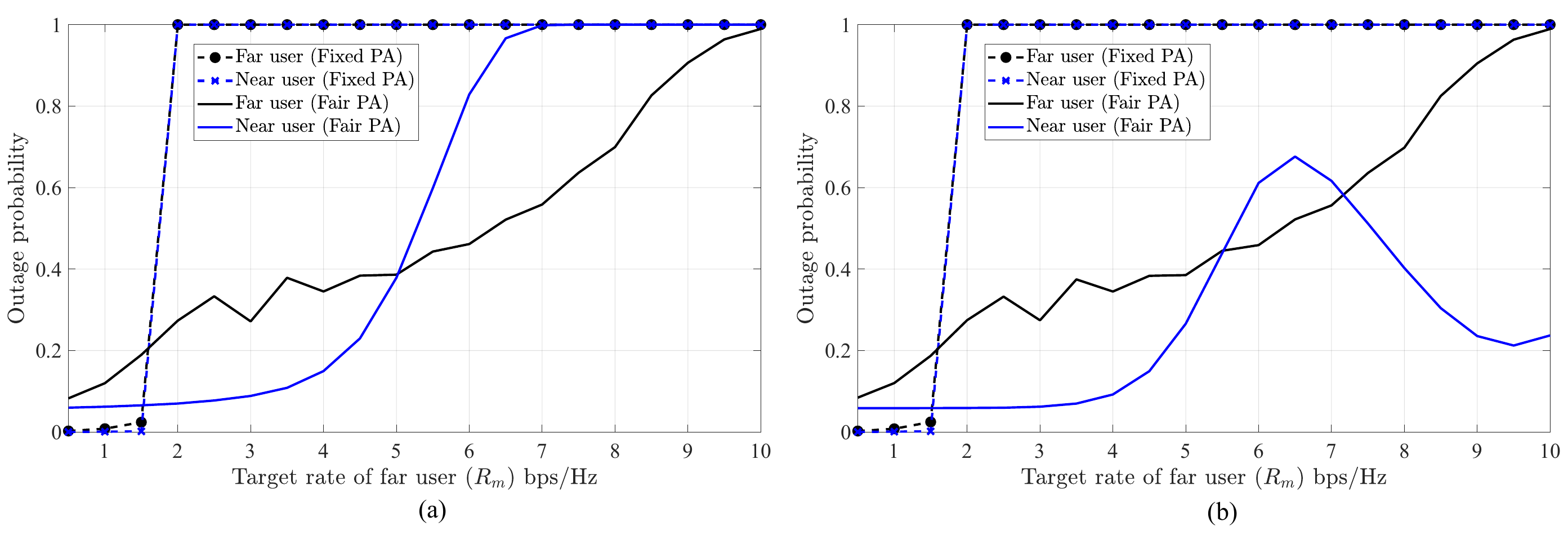}
    \caption{Outage probability comparison of fixed PA and basic fair PA schemes as a function of the target rate for the far user, using (a) Algorithm 1 and (b) Algorithm 2.}
    \label{image4}
\end{figure*}

\begin{algorithm}[!t]
\caption{Basic fair PA scheme (Fair PA)}
\begin{algorithmic}[1]
    \State Initialize: $\alpha_n$, $\alpha_m$, $p_\circ$, $\sigma_\circ^2$, $\boldsymbol{\mathrm{H}}_m$ and $R_m$.
    \State Calculate the target SINR $\xi$ using the equation $\xi = 2^{R_m}-1$.
    \While{${\left\|{\boldsymbol{\mathrm{H}}}_m\right\|}^2 p_\circ \alpha_m - {\left\|{\boldsymbol{\mathrm{H}}}_m\right\|}^2 p_\circ \alpha_n < \sigma_\circ^2$}
        \State Compute $\alpha_m$ using \eqref{GrindEQ27}.
        \If{$\alpha_m > 1$}
            \State Set $\alpha_m = 1$ and $\alpha_n = 0$.
        \Else
            \State Set $\alpha_m = \frac{\xi (p_\circ {\left\|{\boldsymbol{\mathrm{H}}}_m\right\|}^2 + \sigma_\circ^2\boldsymbol{I})}{p_\circ (1+\xi) {\left\|{\boldsymbol{\mathrm{H}}}_m\right\|}^2}$ and $\alpha_n = 1 - \alpha_m$.
        \EndIf
        \State Calculate the rate $R_m$ for the far user using the equation:
        \[\resizebox{0.5\columnwidth}{!}{$
        C_m = \log_2 \left( 1 + \frac{p_\circ \alpha_m {\left\|{\boldsymbol{\mathrm{H}}}_m\right\|}^2}{p_\circ {\left\|{\boldsymbol{\mathrm{H}}}_m\right\|}^2 \alpha_n + \sigma_\circ^2\boldsymbol{I}} \right).
        $}\]
    \EndWhile
    \If{$C_m < R_m$}
        \State Increase the target SINR $\xi$ and go back to step 3.
    \Else
        \State Return the values of $\alpha_n$ and $\alpha_m$ obtained in step 8 as the optimal power allocation coefficients.
    \EndIf
\end{algorithmic}
\end{algorithm}

According to the considered framework, the outage probability of the user $n$ is given as follows
\begin{equation}\resizebox{0.85\columnwidth}{!}{$
\begin{array}{ll}
P^{\mathrm{out}}_n &={\mathrm{Pr} \left\{C_{n\to m}<R_m\ or\ C_n<R_n\right\}\ }\\
          &={\mathrm{1-Pr} \left\{C_{n\to m}\ge R_m\right\}\ }{\mathrm{Pr} \left\{C_n\ge R_n\right\}\ }.\\
\label{GrindEQ23}
\end{array}
$}\end{equation}
Whereas the outage probability of the user $m$ can be expressed as follows
\begin{equation}\resizebox{0.85\columnwidth}{!}{$
P^{\mathrm{out}}_m={\mathrm{Pr} \left\{C_m<R_m\right\}\ }={\mathrm{1-Pr} \left\{C_m\ge R_m\right\}\ }.
\label{GrindEQ25}
$}\end{equation}
The $R_n$ and $R_m$ are the target rates for user $n$ and user $m$, respectively.

\section{Choosing Power Allocation Coefficients}

The scheme from \cite{35} for a SISO system is adapted for MIMO. For simplicity, consider a two-user NOMA framework with user $m$ as the far user and user $n$ as the near user. The goal is to choose ${\alpha }_m$ and ${\alpha }_n$ such that $C_m \ge R_m$. Assuming $C_m = R_m$, \eqref{GrindEQ19} can be rewritten as follows
\begin{equation}\resizebox{0.6\columnwidth}{!}{$
R_m={{\mathrm{log}}_2 \left(1+\frac{{p_{\circ }{\alpha }_m\left\|{\boldsymbol{\mathrm{H}}}_m\right\|}^2}{{p_{\circ }\left\|{\boldsymbol{\mathrm{H}}}_m\right\|}^2{\alpha }_n+{\sigma }^2_{\circ }\boldsymbol{I}}\right).\ }
\label{GrindEQ26}
$}\end{equation}
After making some mathematical manipulations and assumptions i.e., ${\alpha }_n=1-{\alpha }_m$, and $\xi =2^{R_m}-1$, we can rewrite the \eqref{GrindEQ26} as follows
\begin{equation}\resizebox{0.38\columnwidth}{!}{$
{\alpha }_m=\frac{{{\xi (p}_{\circ }\left\|{\boldsymbol{\mathrm{H}}}_m\right\|}^2+{\sigma }^2_{\circ }\boldsymbol{I})}{p_{\circ }(1+\xi ){\left\|{\boldsymbol{\mathrm{H}}}_m\right\|}^2}.
\label{GrindEQ27}
$}\end{equation}
To ensure that ${\alpha }_m$ does not exceed $1$, an upper limit has been imposed on the expression given in \eqref{GrindEQ27}, as follows
\begin{equation}\resizebox{0.6\columnwidth}{!}{$
{\alpha }_m={\mathrm{min} \left(1,\ \frac{{{\xi (p}_{\circ }\left\|{\boldsymbol{\mathrm{H}}}_m\right\|}^2+{\sigma }^2_{\circ }\boldsymbol{I})}{p_{\circ }(1+\xi ){\left\|{\boldsymbol{\mathrm{H}}}_m\right\|}^2}\right)\ }.
\label{GrindEQ28}
$}\end{equation}
Once ${\alpha}_m$ is determined, ${\alpha }_n$ can be computed as ${\alpha }_n=1-{\alpha }_m$. A pseudocode for this basic fair PA scheme is provided in Algorithm 1.

Although better than fixed PA, the basic fair PA scheme has limitations. If the far user has a weak channel with the BS, allocating all power to the far user may still fail to achieve the target rate, causing an outage. Setting $\alpha_m=1$ (and $\alpha_n=0$) can also result in the near user being in outage. To address this, a modification is proposed (see Algorithm 2). When $\frac{{{\xi (p}_{\circ }\left|{\boldsymbol{\mathrm{H}}}m\right|}^2+{\sigma }^2_{\circ }\boldsymbol{I})}{p_{\circ }(1+\xi ){\left|{\boldsymbol{\mathrm{H}}}_m\right|}^2}>1$, the scheme sets ${\alpha }_m=0$ and ${\alpha }_n=1$. This ensures efficient power utilization by avoiding allocation to users unable to meet rate targets.

\begin{algorithm}
\caption{Modified Basic Fair PA (Improved Fair PA)}
\begin{algorithmic}[1]
    \State Initialize: $\alpha_n$, $\alpha_m$, $p_\circ$, $\sigma_\circ^2$, $\boldsymbol{\mathrm{H}}_m$, and $R_m$.
    \State Calculate the target SINR $\xi$ using the equation $\xi = 2^{R_m} - 1$.
    \While{${\left\|{\boldsymbol{\mathrm{H}}}_m\right\|}^2 p_\circ \alpha_m - {\left\|{\boldsymbol{\mathrm{H}}}_m\right\|}^2 p_\circ \alpha_n < \sigma_\circ^2$}
        \State Compute $\alpha_m$ using \eqref{GrindEQ27}.
        \If{$\xi(p_\circ {\left\|{\boldsymbol{\mathrm{H}}}_m\right\|}^2 + \sigma_\circ^2\boldsymbol{I}) \leq p_\circ (1 + \xi) {\left\|{\boldsymbol{\mathrm{H}}}_m\right\|}^2$}
            \State $\alpha_m = \frac{\xi(p_\circ {\left\|{\boldsymbol{\mathrm{H}}}_m\right\|}^2 + \sigma_\circ^2\boldsymbol{I})}{p_\circ (1 + \xi) {\left\|{\boldsymbol{\mathrm{H}}}_m\right\|}^2}$.
        \Else
            \State Set $\alpha_m = 0$ and $\alpha_n = 1$.
        \EndIf
        \State Calculate the rate $R_m$ for the far user using the equation:
        \[\resizebox{0.5\columnwidth}{!}{$
        R_m = \log_2 \left( 1 + \frac{p_\circ \alpha_m {\left\|{\boldsymbol{\mathrm{H}}}_m\right\|}^2}{p_\circ {\left\|{\boldsymbol{\mathrm{H}}}_m\right\|}^2 \alpha_n + \sigma_\circ^2\boldsymbol{I}} \right).
        $}\]
    \EndWhile
    \If{$C_m < R_m$}
        \State Increase the target SINR $\xi$ and go back to step 3.
    \Else
        \State Return the values of $\alpha_n$ and $\alpha_m$ obtained in step 3 as the optimal power allocation coefficients.
    \EndIf
\end{algorithmic}
\end{algorithm}

\section{Results and Conclusion}

This section evaluates the considered TeraRIS NOMA communications framework for a $16\times 16$ MIMO system. The results are detailed in the following subsections. Existing THz simulators from \cite{2,52} were modified for simulations. The settings are: operating frequency $0.3$ THz\footnote{An operating frequency of $0.3$ THz was selected based on its known stability and reduced atmospheric gas-induced attenuation fluctuations as reported in \cite{53}.}, $200$ RIS elements, $16$ antennas at BS and user, user $m$ and $n$ distances from BS are $500$ m and $250$ m, respectively, user $m$ and $n$ distances from RIS are $150$ m and $250$ m, respectively, transmitted power of $30$ dB, and RIS is $100$ m from BS.

\subsection{Outage Analysis}

This subsection presents the results obtained by plotting the outage probability against the far user's target rate $R_m$, with an overall transmit power of $30$ dBm. The same target rate was used to calculate the outage probability for both users.

\begin{figure}[t]
    \centering
    \includegraphics[width=0.9\columnwidth]{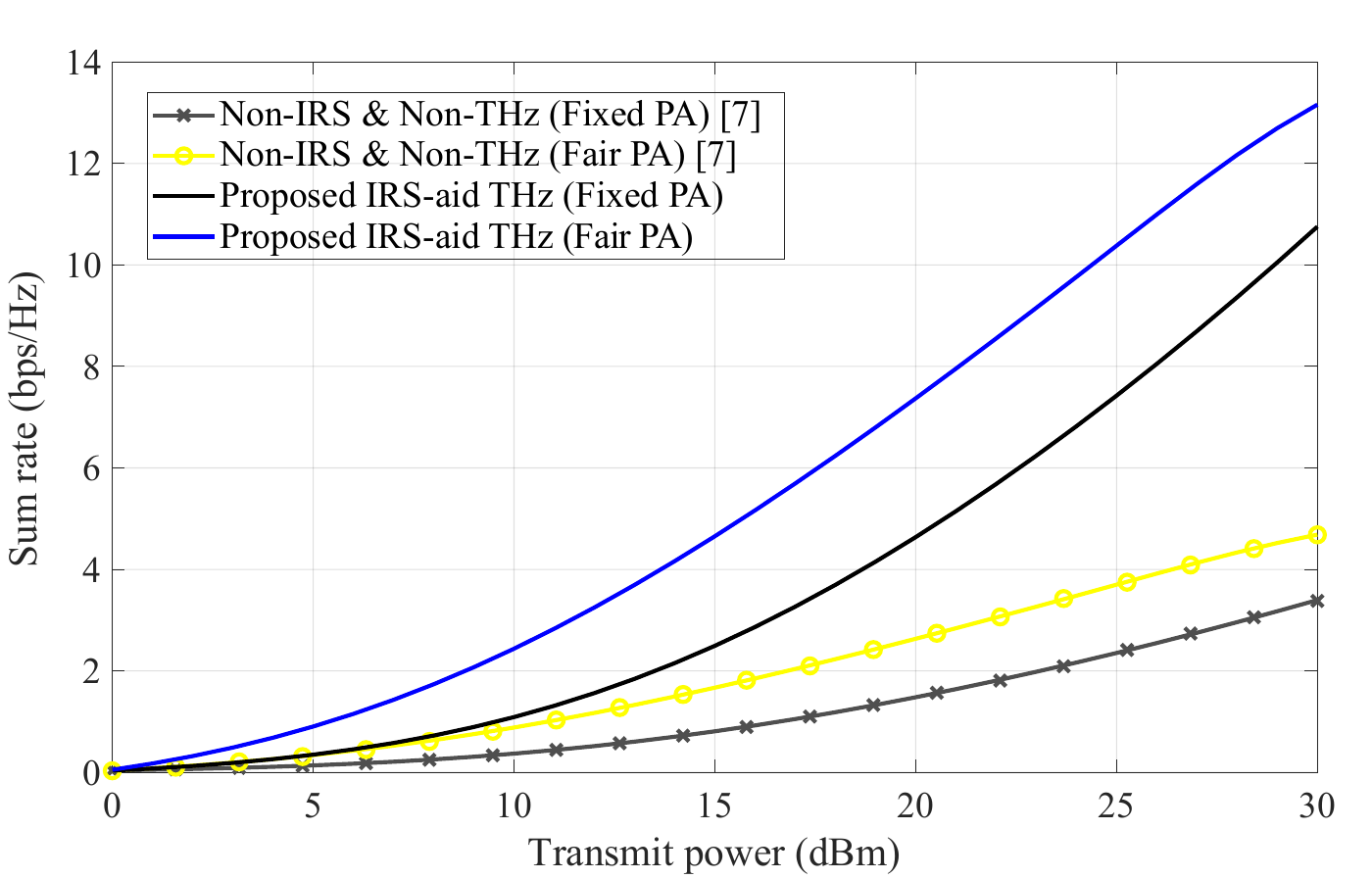}
    \caption{Comparison of sum rates between Non-RIS Non-THz networks and the proposed network for fixed PA and fair PA schemes, concerning transmitted power.}
    \label{image6}
\end{figure}

Fig.~\ref{image4}(a) shows that the fixed PA scheme performs poorly due to constant power allocation, resulting in a high outage probability when the far user's channel deteriorates or target rate increases. The basic fair PA scheme adjusts power allocation coefficients ${\alpha }_m$ and ${\alpha }_n$ based on instantaneous CSI and target rates, reducing outage probability, especially for the far user. However, the near user's outage probability increases sharply and then plateaus as $R_m$ increases, reflecting a trade-off between fairness and efficiency. The improved fair PA scheme optimizes power allocation, setting ${\alpha }_m=0$ when needed to prioritize the near user's performance. Fig.~\ref{image4}(b) shows that the far user's outage remains similar to the basic fair PA scheme, while the near user's outage probability initially increases, peaks, and then decreases with balanced power allocation at higher $R_m$. At a specific target rate (e.g., $R_m = 5$ bps/Hz), Algorithm 2 performs approximately $35.7$\% better than Algorithm 1. This scheme balances fairness and efficiency, providing a robust solution for wireless communications systems.

\subsection{Capacity Analysis}

Fig.~\ref{image6} compares the sum rate ($R_n+R_m$) achieved by the improved fair PA, fixed PA algorithms, and the non-RIS non-THz framework from \cite{35}. The proposed scheme achieves up to a $23$ and $184$\% sum-rate gain over proposed RIS-aid THz fixed PA and non-RIS and non-THz fair PA \cite{35}, respectively, at $30$ dBm. The sum rate demonstrates the superiority of the fair PA algorithm over the fixed PA and non-RIS non-THz schemes across different transmit power levels. This advantage comes from the fair PA algorithm's dynamic power allocation, which adapts to changing channel conditions and user demands. In contrast, the fixed PA scheme's static allocation results in suboptimal performance, and the non-RIS non-THz framework emphasizes the gains from RIS-assisted THz communications. 

\section{Conclusion}
This paper presents a framework for 6G and beyond, leveraging RIS to enhance the performance of NOMA-MIMO systems for THz band industrial communications. It derives the closed-form capacity expressions and addresses power allocation for near and far users for the considered framework. Fixed PA schemes fail to adapt to dynamic channels, whereas fair PA schemes adjust power allocation based on channel conditions, improving sum rates and reducing outages. This framework enhances spectral efficiency, ensures fair resource distribution, and mitigates path losses, improving system capacity. For practical deployment, hardware constraints such as RIS phase quantization, THz power amplifier nonlinearities, and synchronization challenges in URLLC must be addressed. Future research should incorporate hardware-aware design, nonlinearity mitigation, and robust synchronization mechanisms to bridge the gap between theory and real-world 6G industrial systems.



\ifCLASSOPTIONcaptionsoff
  \newpage
\fi
\bibliographystyle{IEEEtran}
\bibliography{Reference}

\end{document}